\newcommand{\ket}[1]{|{#1}\rangle}
\def\one{{\mathchoice {\rm 1\mskip-4mu l} {\rm 1\mskip-4mu l} {\rm
1\mskip-4.5mu l} {\rm 1\mskip-5mu l}}}

\documentclass[a4paper]{spie} 
\usepackage[]{graphicx}

\title{Quantum Simulations of Physics Problems}

\author{Rolando Somma, Gerardo Ortiz, Emanuel Knill, and James Gubernatis\\ Los
Alamos National Laboratory, Los Alamos, USA}

\begin{document} \maketitle

\begin{abstract}
If a large Quantum Computer (QC) existed today, what type of physical
problems could we efficiently simulate on it that we could not simulate
on a classical Turing machine? In this paper we argue that a QC could
solve some relevant physical ``questions" more efficiently. The
existence of one-to-one mappings between different algebras of
observables or between different Hilbert spaces allow us to represent
and imitate any physical system by any other one (e.g., a bosonic
system by a spin-1/2 system). We explain how these mappings can be
performed showing quantum networks useful for the efficient evaluation
of some physical properties, such as correlation functions and energy
spectra.
\end{abstract}



\section{introduction}
\label{section1}

Quantum simulation of physical systems on a QC has acquired
importance during the last years since it is believed that QCs can
simulate quantum physics problems more efficiently than their classical
analogues \cite{feynman1982}: The number of operations needed for 
deterministically solving a quantum many-body problem on a classical
computer (CC) increases exponentially with the number of degrees of
freedom of the system. 

In quantum mechanics, each physical system has associated a language of
operators and an algebra realizing this language, and can be considered
as a possible model of quantum computation \cite{ortiz2001}. As we
discussed in a previous paper \cite{somma2002}, the existence of
one-to-one mappings between different languages (e.g., the
Jordan-Wigner transformation that maps fermionic operators onto spin-1/2
operators) and between quantum states of different Hilbert spaces,
allows the quantum simulation of one physical system by any other one.
For example, a liquid nuclear magnetic resonance QC (NMR) can simulate
a system of $^4$He atoms (hard-core bosons) because an isomorphic
mapping between both algebras of observables exists.

The existence of mappings between operators allows us to construct
quantum network models from sets of elementary gates, to which we map
the operators of our physical system. An important remark is that these
mappings can be performed efficiently: we need a number of steps that
scales polynomially with the system size. However, this fact alone is not
sufficient to establish that any quantum problem can be solved
efficiently. One needs to show that all steps involved in the
simulation (i.e., preparation of the initial state, evolution,
measurement, and measurement control) can be performed with polynomial
complexity. For example, 
the number of different eigenvalues in the two-dimensional
Hubbard model scales exponentially with the system size, so
QC algorithms for obtaining its energy spectrum will also require
a number of
operations that scales exponentially with the system size 
\cite{somma2002}.

Typically, the degrees of freedom of the physical system over which we
have quantum control constitute the model of computation. In this
paper, we consider the simulation of any physical system by the
standard model of quantum computation (spin-1/2 system), since this
might be the language needed for the  practical implementation of the
quantum algorithms (e.g., NMR). Therefore, the complexity of the 
quantum algorithms is analyzed from the point of view of the number of
resources (elementary gates) needed for their implementation in the
language of the standard model. Had another model of computation being
used, one should follow the same qualitative steps although the
mappings and network structure would be different.  

The main purpose of this work is to show how to simulate any physical
process and system using the least possible number of resources. We
organized the paper in the following way: In section \ref{section2} we
describe the standard model of quantum computation (spin-1/2 system).
Section \ref{section3} shows the mappings between physical systems
governed by a generalized Pauli's exclusion principle (fermions, etc.)
and the standard model, giving examples of algorithms for the first two
steps (preparation of the initial state and evolution) of the quantum
simulation. In section \ref{section4} we develop similar steps for the
simulation of quantum systems whose language has an
infinite-dimensional representation, thus, there is no exclusion
principle (e.g., canonical bosons). In section \ref{section5} we
explain the measurement process used to extract information of some
relevant and generic physical properties, such as correlation functions
and energy spectra. We conclude with a discussion about efficiency and
quantum errors (section \ref{section6}), and a summary about the
general statements (section \ref{section7}).


\section{standard model of quantum computation}
\label{section2}

In the standard model of quantum computation, the fundamental unit is
the {\it qubit}, represented by a two level quantum system $\ket{\sf a}
= a \ket{0} + b\ket{1}$. For a spin-1/2 particle, for example, the two
``levels'' are the two different orientations of the spin,
$\ket{\uparrow}=\ket{0}$ and $\ket{\downarrow}=\ket{1}$. In this
model, the algebra assigned to a system of $N$-qubits is built upon the
Pauli spin-1/2 operators $\sigma_x^j$, $\sigma_y^j$ and $\sigma_z^j$
acting on the $j$-th qubit (individual qubit).  The commutation
relations for these operators satisfy an $\bigoplus\limits_{i=1}^N
su(2)_i$ algebra defined by ($\mu,\nu,\lambda=x,y,z$)
\begin{equation}
\label{su2}
[\sigma_{\mu}^j,\sigma_{\nu}^k]=2i\delta_{jk}\epsilon_{\mu \nu \lambda} 
\sigma_{\lambda}^j ,
\end{equation}
where $\epsilon_{\mu \nu \lambda}$ is the totally anti-symmetric
Levi-Civita symbol. Sometimes it is useful to write the commutation
relations in terms of the raising and lowering spin-1/2 operators 
\begin{equation}
\label{sigmapm}
\sigma_{\pm}^j = \frac{\sigma_x^j \pm i \sigma_y^j}{2}.
\end{equation}

Any operation on a QC is represented by a unitary operator $U$ that
evolves some initial state (boot-up state) in a way that satisfies the
time-dependent Schr\"odinger equation for some Hamiltonian $H$. Any
unitary operation (evolution) $U$ applied to a system of $N$ qubits can
be decomposed into either single qubit rotations $R_{\mu}(\vartheta)=
e^{-i \frac {\vartheta}{2} \sigma_{\mu}}$ by an angle $\vartheta$ about
the $\mu$ axis or two qubits Ising interactions $R_{z^j, z^k}=e^{i
\omega \sigma_z^i \sigma_z^j}$. This is an important result of quantum
information, since with these operations one can perform universal
quantum computation. It is important to mention that we could also
perform universal quantum computation with single qubit rotations and
C-NOT gates \cite{nielsen2000} or even with different control
Hamiltonians. The crucial point is that we need to have quantum control
over those elementary operations in the real physical system.  

In the following, we will write down our algorithms in terms of single
qubit  rotations and two qubits Ising interactions, since this is the
language needed for the implementation of the algorithms, for example,
in a  liquid NMR QC. Again, had we used a different set of elementary
gates our main results still hold but with modified quantum networks. 

As an example of such decompositions, we consider the unitary operator
$U(t)=e^{iHt}$, where $H=\alpha \sigma_x^1 \sigma_z^2 \sigma_x^3$
represents a time-independent Hamiltonian. After some simple
calculations \cite{ortiz2001,somma2002} we decompose $U$ into
elementary gates (one qubit rotations and two qubits interactions) in
the following way
\begin{equation}
\label{decomp1}
U(t)=e^{i \alpha \sigma_x^1 \sigma_z^2 \sigma_x^3 t} =  
e^{-i \frac{\pi}{4} \sigma_y^3}
e^{i \frac{\pi}{4} \sigma_z^1\sigma_z^3} e^{i \frac{\pi}{4} \sigma_x^1}
e^{i \alpha \sigma_z^1 \sigma_z^2 t} e^{-i \frac{\pi}{4} \sigma_x^1}
e^{-i \frac{\pi}{4} \sigma_z^1\sigma_z^3} e^{i \frac{\pi}{4} \sigma_y^3}.
\end{equation}

This decomposition is shown in Fig. 1, where the quantum network
representation is displayed. In the same way, we could also decompose
an operator $U'(t)=e^{-i\alpha\sigma_y^1 \sigma_z^2 \sigma_y^3 t}$
using similar steps, by replacing $\sigma_x^i  \leftrightarrow
\sigma_y^i$ in the right hand side of Eq. \ref{decomp1}.
\begin{figure}[hbt]
\begin{center}
\includegraphics[width=9.8cm]{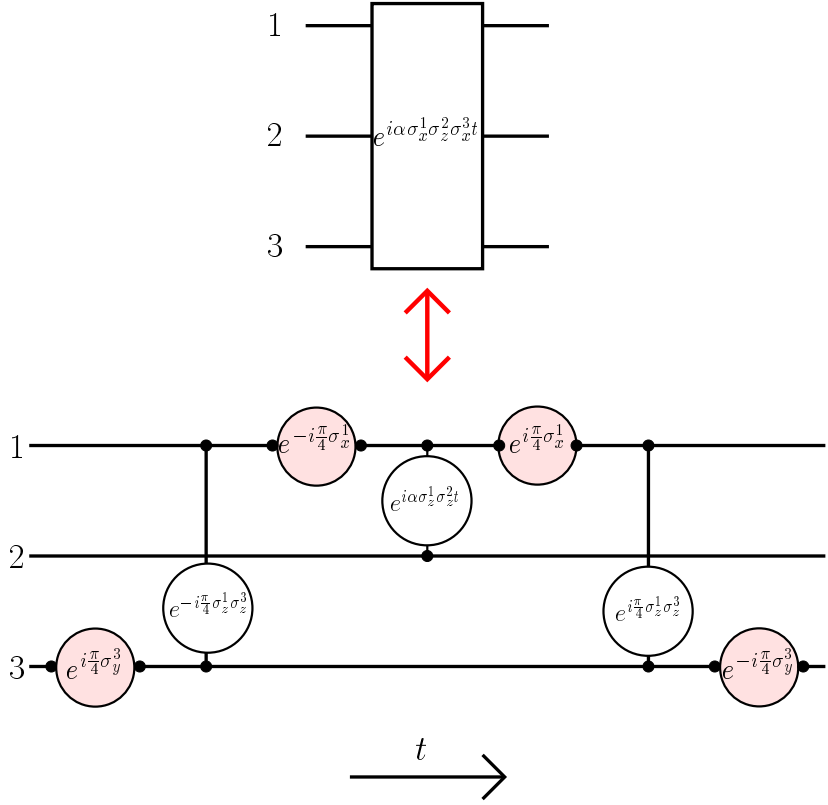}
\end{center}
\caption{Decomposition of the unitary operator $U(t)=e^{i \alpha
\sigma_x^1 \sigma_z^2 \sigma_x^3 t}$ into elementary single qubit
rotations and two qubits interactions. Time $t$ increases from left to
right.}
\label{fig:1}
\end{figure}

\section{simulation of fermionic systems}
\label{section3}

As discussed in the Introduction, quantum simulations require
simulations of systems with diverse degrees of freedom and particle
statistics. Fermionic systems are governed by Pauli's exclusion
principle, which implies that no more than one fermion can occupy the
same quantum state at the same time. In this way, the Hilbert space of
quantum states that represent a system of fermions in a solid is
finite-dimensional ($2^N$ for spinless fermions, where $N$ is the
number of sites or modes in the solid), and one could think in the
existence of one-to-one mappings between the fermionic and Pauli's
spin-1/2 algebras. Similarly, any language which involves operators
with a finite-dimensional representation (e.g., hard-core bosons, higher
irreps of $su(2)$, etc.) can be mapped onto the standard model language
\cite{review}. 

In the second quantization representation, the (spinless) fermionic
operators $c^{\dagger}_i$ ($c^{\;}_i$) are defined as the creation
(annihilation) operators of a fermion in the $i$-th mode
($i=1,\cdots,N$). Due to the Pauli's exclusion principle and the
antisymmetric nature of the  fermionic wave function under the
permutation of two fermions, the fermionic algebra is given by the
following commutation relations
\begin{equation}
\label{fermcom}
\{ c_i,c_j \}=0 ,  \mbox{ } \{ c^{\dagger}_i,c_j \} = \delta_{ij}
\end{equation}
where $\{,\}$ denotes the anticommutator.

The Jordan-Wigner transformation \cite{jordan1928} is the isomorphic
mapping that allows the description of a fermionic system by the
standard model
\begin{eqnarray}
\label{JW}
c_j \rightarrow \left( \prod\limits_{l=1}^{j-1} -\sigma_z^l \right) \sigma_-^j\\
\label{JW2}
c^{\dagger}_j \rightarrow \left( \prod\limits_{l=1}^{j-1} -\sigma_z^l \right) 
\sigma_+^j ,
\end{eqnarray}
where $\sigma_{\mu}^i$ are the Pauli operators defined in section 
\ref{section2}.
One can easily verify that if the operators $\sigma_{\mu}^i$ satisfy
the  $su(2)$ commutation relations (Eq. \ref{su2}), the operators
$c^{\dagger}_i$ and $c^{\;}_i$ obey Eqs. \ref{fermcom}.

We now need to show how to simulate a fermionic system by a QC. Just as
for a simulation on a CC, the quantum simulation has three basic 
steps: the preparation of an initial state, the evolution of this
state, and the measurement of a relevant physical property of the
evolved state. We will now explain the first two steps, postponing the
third until section \ref{section5}.

\subsection{Preparation of the initial state}
\label{section3.1}

In the most general case, any quantum state $\ket{\psi}$  of $N_e$
fermions can be written as a linear combination of Slater determinants
$\ket{\phi_\alpha}$
\begin{equation}
\ket{\psi}=\sum\limits_{\alpha=1}^L g_\alpha \ \ket{\phi_\alpha} , 
\end{equation}
where
\begin{equation} 
\label{slater1}
\ket{\phi_\alpha} = \prod \limits_{j=1}^{N_e} c^{\dagger}_j \ \ket{\sf vac}
\end{equation}
with the vacuum state $\ket{\sf vac}$ defined as the state with no
fermions. In the spin language, $\ket{\sf vac}=\ket{\downarrow
\downarrow \cdots \downarrow}$.

We can easily prepare the states $\ket{\phi_\alpha}$ by noticing that
the quantum gate, represented by the unitary operator
\begin{equation}
\label{Um}
U_m=e^{i \frac{\pi}{2} (c^{\;}_m+c^\dagger_m)}
\end{equation}
when acting on the vacuum state, produces $c^\dagger_m \ket{0}$ up to a
phase factor. Making use of the Jordan-Wigner transformation (Eqs.
\ref{JW}, \ref{JW2}), we can write the operators $U_m$ in the spin
language
\begin{equation}
U_m=e^{i \frac{\pi}{2} \sigma_x^m \prod\limits_{j=1}^{m-1} -\sigma_z^j}.
\end{equation}
The successive application of $N_e$ similar unitary operators will
generate the state $\ket{\phi_\alpha}$ up to an irrelevant global
phase.

A detailed preparation of the fermionic state $\ket{\psi}=\sum
\limits_{\alpha=1}^L g_\alpha \ \ket{\phi_\alpha}$ can be found in a
previous work \cite{ortiz2001}. The  basic idea is to use $L$ extra
(ancilla) qubits, then perform unitary evolutions controlled in the
state of the ancillas, and finally perform a measurement of the
$z$-component of the spin of the ancillas. In this way, the probability
of successful preparation of $\ket{\psi}$ is $1/L$. (We need of the
order of $L$ trials before a successful preparation.)

Another important case is the preparation of a Slater
determinant in a different basis than the one given before
\begin{equation}
\label{slater2}
\ket{\phi_\beta}=\prod\limits_{i=1}^{N_e} d^\dagger_i \ \ket{\sf vac} ,
\end{equation}
where the fermionic operators $d^\dagger_i$'s are related to the operators
$c^\dagger_j$ through the following canonical transformation
\begin{equation}
\label{unitmap}
\overrightarrow{d}^\dagger = e^{iM} \overrightarrow{c}^\dagger
\end{equation}
with
$\overrightarrow{d}^\dagger=(d^\dagger_1,d^\dagger_2,\cdots,d^\dagger_N)$,
$\overrightarrow{c}^\dagger=(c^\dagger_1,c^\dagger_2,\cdots,c^\dagger_N)$,
and $M$ is an $N \times N$ Hermitian matrix. Making use of Thouless's
theorem \cite{blaizot1986}, we observe that one Slater determinant
evolves into the other, $\ket{\phi_\beta}=U\ket{\phi_\alpha}$, where
the unitary operator $U=e^{-i \overrightarrow{c}^\dagger M
\overrightarrow{c}}$ can be written in spin operators using the
Jordan-Wigner transformation and can be decomposed into elementary gates
\cite{somma2002}, as described in section \ref{section2}. Since the
number of gates scales polynomially with the system size,  the state
$\ket{\phi_\beta}$ can be  efficiently prepared from the state
$\ket{\phi_\alpha}$.

\subsection{Evolution of the initial state}
\label{section3.2}

The second step in the quantum simulation is the evolution of the
initial state. The unitary evolution operator of a time-independent
Hamiltonian $H$ is $U(t)= e^{iHt}$. In general, $H=K+V$ with $K$
representing the kinetic energy and $V$ the potential energy. Since we
usually have $[K,V] \neq 0$, the decomposition of $U(t)$, written in
the spin language through the Jordan-Wigner transformation (Eqs.
\ref{JW},\ref{JW2}), in terms of elementary gates (one qubit rotations
and two qubits interactions), becomes complicated. To avoid this
problem, we instead use a Trotter decomposition, so the evolution
during a short period of time  ($\Delta t=t/{\cal N}$ with $\Delta
t\rightarrow 0$) is approximated. To order ${\sf O}(\Delta t)$ (first
order Trotter breakup)
\begin{eqnarray}
\label{trotter}
U(t)&=&\prod\limits_{g=1}^{\cal N} U(\Delta t), \\
U(\Delta t)&=& e^{i H \Delta t}= e^{ i (K+V) \Delta t} \sim  e^{i K \Delta t}
e^{i V \Delta t}.
\end{eqnarray}

The potential energy $V$ is usually a sum of commuting diagonal terms,
and the decomposition of $e^{i V \Delta t}$ into elementary gates is
straightforward. However, the kinetic energy $K$  is usually a sum of
noncommuting terms of the form  $c^\dagger_i c^{\;}_j + c^\dagger_j
c^{\;}_i$ (bilinear fermionic operators), so we need again to perform a
Trotter approximation of the operator $e^{i K \Delta t}$. As an example
of such a decomposition, we consider a typical term  $e^{i (c^\dagger_i
c^{\;}_j +c^\dagger_j c^{\;}_i)\Delta t}$ ($i<j$), when mapped onto the
spin language gives
\begin{equation}
\label{decomp2}
e^{-\frac{i}{2}  (\sigma_x^i \sigma_x^j +\sigma_y^i \sigma_y^j) 
 \prod\limits_{k=
i+1}^{j-1}(-\sigma_z^k) } = e^{-\frac{i}{2}  
\sigma_x^i \sigma_x^j 
 \prod\limits_{k=
i+1}^{j-1}(-\sigma_z^k) } e^{-\frac{i}{2}  
\sigma_y^i \sigma_y^j
\prod\limits_{k=
i+1}^{j-1}(-\sigma_z^k) }.
\end{equation}
 
The decomposition of each term on the right hand side of Eq. 
\ref{decomp2} into  elementary gates was already described in  previous
work \cite{somma2002}. In section \ref{section2} and Fig. 1, we also
showed an example of such a decomposition for $i=1$ and $j=3$. It is
important to mention that the required number of elementary gates
scales polynomially with the length $|j-i|$. Notice that this step is
not necessary for bosonic systems since no string of $\sigma_z^k$
operators is involved (see section \ref{section4}).

The accuracy of this method increases as $\Delta t$ decreases, so we
might  require a large number of gates to perform the evolution with
small errors. To overcome this problem, one could use Trotter
approximations of higher order in $\Delta t$ \cite{suzuki1993}.

\subsection{Generalization: simulation of anyonic systems}
\label{section3.3}

The concepts described in sections \ref{section3.1} and
\ref{section3.2} can be easily generalized to other more general
particle statistics, namely hard-core anyons. By ``hard-core", we mean
that only zero or one particle can occupy a single mode (Pauli's exclusion
principle). 

The commutation relations between the anyonic creation and annihilation
operators $a^{\dagger}_i$ and $a_i$, are  given by 
\begin{eqnarray}
\label{anyoncom1}
[ a^{\;}_i,a^{\;}_j ]_\theta &=& 
[ a^\dagger_i,a^\dagger_j ]_\theta=0  
 \ , \nonumber\\
{[}a^{\;}_i,a^\dagger_j{]}_{-\theta}&=&\delta_{ij} (1-(e^{-i
\theta}+1)n_j) \ , \\
{[} n_i, a^\dagger_j ]&=& \delta_{ij} 
a^\dagger_j  \ , \nonumber
\end{eqnarray}
($i \leq j$) where $n_j=a^\dagger_j a^{\;}_j$,  $[\hat{A},\hat{B}
]_\theta = \hat{A} \hat{B} - e^{i \theta} \hat{B} \hat{A}$,  with $0
\leq \theta < 2\pi$ defining the statistical angle. In particular, 
$\theta=\pi$  mod($2\pi$) corresponds to canonical spinless fermions,
while $\theta=0$ mod($2\pi$) represents hard-core bosons. 

In order to simulate this problem with a QC made of qubits, we  need to
apply the following isomorphic and efficient mapping between algebras
\begin{eqnarray}
a^{\dagger}_j &=& \prod\limits_{i<j} [\frac {e^{-i \theta} +1}{2} + 
\frac {e^{-i \theta} -1}{2} \sigma_z^i] \ \sigma_+^j , \nonumber \\
a_j &=& \prod\limits_{i<j} [\frac {e^{i \theta} +1}{2} + 
\frac {e^{i \theta} -1}{2} \sigma_z^i] \ \sigma_-^j ,  \\
n_j &=& \frac{1}{2} (1+ \sigma_z^j) ,\nonumber
\end{eqnarray}
where the Pauli operators $\sigma_{\mu}^j$ where defined in section 
\ref{section2}, and  since they satisfy Eq. \ref{su2}, the
corresponding commutation relations for the anyonic operators (Eqs.
\ref{anyoncom1}) are satisfied, too.

We can now proceed in the same way as in the fermionic case, writing
our anyonic evolution operator in terms of single qubit rotations and
two qubits interactions in the spin-1/2 language. As we already
mentioned, anyon statistics have fermion and hard-core boson
statistics as limiting cases. We now relax the hard-core condition on
the bosons. 

\section{Simulation of Bosonic systems} 
\label{section4}

Quantum computation is based on the manipulation of quantum systems
that possess finite number of degrees of freedom (e.g., qubits). From
this point of view, the simulation of bosonic systems appears to be
impossible, since the non existence of an exclusion principle implies
that the Hilbert space used to represent bosonic quantum sates is
infinite-dimensional; that is, there is no limit to the number of
bosons that can occupy a given mode. However, sometimes we might be
interested in simulating and studying properties such that the use of
the whole Hilbert space is unnecessary, and only a finite  sub-basis of
states is sufficient.  This is the case for physical systems with
interactions given by the  Hamiltonian
\begin{equation}
\label{bosonhamilt}
H= \sum \limits_{i,j=1}^N \alpha_{ij} \ b^{\dagger}_i b^{\;}_j +
\beta_{ij} \ n_i n_j ,
\end{equation}
where the operators $b^{\dagger}_i$ ($b^{\;}_i$) create (destroy) a
boson  at site $i$, and $n_i=b^{\dagger}_i b^{\;}_i$ is the number
operator. The space dimension of the lattice is encoded in the
parameters $\alpha_{ij}$ and $\beta_{ij}$. Obviously, the total number
of bosons $N_P$ in the system is conserved, and we restrict ourselves
to work with a finite sub-basis of states, where the dimension depends
on the value of $N_P$. 

The respective bosonic commutation relations (in an
infinite-dimensional Hilbert space) are
\begin{equation}
\label{bosoncom}
[b_i,b_j]=0 , [b_i,b^{\dagger}_j]=\delta_{ij}.
\end{equation}
However, in a finite basis of states represented by $\{
\ket{n_1,n_2,\cdots,n_N }$ with $n_i=0,\cdots, N_P \}$, where $N_P$ is
the maximum number of bosons per site, the operators $b^{\dagger}_i$
can have the following matrix  representation
\begin{equation}
\label{bosonprod}
\bar{b}^{\dagger}_i =\one \otimes \cdots \otimes \one \otimes 
\underbrace{\hat {b}^{\dagger}}_{i^{th}\mbox{ factor}} \otimes
\one \otimes \cdots \otimes \one \\
\end{equation}
where $\otimes$ indicates the usual tensorial product between matrices,
and the $(N_P+1) \times (N_P+1)$ dimensional matrices $\one$ and
$\hat{b}^{\dagger}$  are
\begin{equation}
\label{bosonrep}
\one =\pmatrix {1&0&0&\cdots&0 \cr  0&1&0&\cdots&0\cr 0&0&1&\cdots&0 \cr
\vdots&\vdots&\vdots&\cdots &\vdots\cr 0&0&0&\cdots&1} 
\mbox{ , }
\hat{b}^{\dagger} = \pmatrix{0 & 0 & 0 &\cdots &0 & 0 \cr 1 & 0 & 0 & 
\cdots & 0& 0 \cr 0& \sqrt{2} & 0 &\cdots &0 & 0 \cr \vdots & \vdots & \vdots
&\cdots &\vdots &\vdots \cr 0 & 0 & 0 &\cdots &\sqrt{N_P} & 0}.
\end{equation}
It is important to note that in this finite basis, the commutation
relations of the bosons $\bar{b}^{\dagger}_i$ differ from the standard
bosonic ones (Eq. \ref{bosoncom}) \cite{review} 
\begin{equation}
\label{bosoncom2}
[\bar{b}^{\;}_i,\bar{b}^{\;}_j]=0 , \mbox{ } 
[\bar{b}^{\;}_i,\bar{b}^{\dagger}_j]=\delta_{ij} \left[ 1-
\frac{N_P+1}{N_P!} (\bar{b}^{\dagger}_i)^{N_P}(\bar{b}^{\;}_i)^{N_P}
\right] ,
\end{equation}
and clearly $(\bar{b}^{\dagger}_i)^{N_P+1}=0$.

As we mentioned in the Introduction, our idea is to simulate any
physical  system in a QC made of qubits. For this purpose, we need to
map the bosonic algebra into the spin-1/2 language. However, since Eqs.
\ref{bosoncom2}  imply that the linear span of the 
operators  $\bar{b}^{\dagger}_i$ and
$\bar{b}^{\;}_i$ is not closed under the bracket (commutator), 
a direct mapping between the bosonic algebra
and the spin-1/2 algebra (such as the case of the  Jordan-Wigner
transformation between the fermionic and spin-1/2 algebra) is not 
possible. Therefore, we could think in a one-to-one mapping between
the  bosonic and spin-1/2 quantum states, instead of an isomorphic
mapping between algebras. Let us show a possible mapping of quantum
states.

We start by considering only the $i$-th site in the chain. Since this
site can be occupied with at most $N_P$ bosons, it is possible to
associate an $N_P+1$ qubits quantum state to each particle number
state, in the following way
\begin{eqnarray}
\label{bosonmap}
|0\rangle_i &\leftrightarrow& |\uparrow_0 \downarrow_1 \downarrow_2 \cdots 
\downarrow_{N_P} \rangle_i \nonumber \\
|1\rangle_i &\leftrightarrow& |\downarrow_0 \uparrow_1 \downarrow_2 \cdots 
\downarrow_{N_P} \rangle_i \nonumber \\
|2\rangle_i &\leftrightarrow& |\downarrow_0 \downarrow_1 \uparrow_2 \cdots 
\downarrow_{N_P} \rangle_i  \\
\vdots && \vdots \nonumber \\
|N_P\rangle_i &\leftrightarrow& |\downarrow_0 \downarrow_1 \downarrow_2
\cdots \uparrow_{N_P} \rangle_i \nonumber
\end{eqnarray} 
where $\ket{n}_i$ denotes a quantum state with $n$ bosons in site $i$.
Therefore, we need $N(N_P+1)$ qubits for the simulation (where $N$
is the number of sites). In Fig. 2
we show an example of this mapping for a quantum state with 7 bosons in a
chain of 5 sites.

By definition (see Eqs. \ref{bosonprod}, \ref{bosonrep}) 
$\bar{b}^{\dagger}_i \ \ket{n}_i= \sqrt{n+1} \ \ket{n+1}_i$, so the
operator 
\begin{equation}
\label{bosonmap2}
\bar{b}^{\dagger}_i = \sum \limits
_{n=0}^{N_P-1} \sqrt{n+1} \ \sigma_-^{n,i} \sigma_+^{n+1,i} ,
\end{equation}
where the pair $(n,i)$ indicates the qubit $n$ that represents the
$i$-th site, acts in the $N_P+1$ qubits states of Eqs. \ref{bosonmap}
as $\bar{b}^{\dagger}_i \ket{ \downarrow_0 \cdots  \downarrow_{n-1} 
\uparrow_n \downarrow_{n+1} \cdots \downarrow_{N_P} }_i = \sqrt{n+1} \ 
\ket{ \downarrow_0 \cdots \downarrow_n \uparrow_{n+1} \downarrow_{n+2}
\cdots \downarrow_{N_P} }_i $. Then, its matrix representation in this
basis is the same matrix representation of $b^{\dagger}_i$ in the
basis of bosonic states. Similarly, the number operator can be written
\begin{equation}
\label{bosonmap2d}
\bar{n}_i = \sum \limits _{n=0}^{N_P} n \ \frac{\sigma_z^{n,i}+1}{2} ,
\end{equation}
and act as $\bar{n}_i \ket{ \downarrow_0 \cdots  \downarrow_{n-1} 
\uparrow_n \downarrow_{n+1} \cdots \downarrow_{N_P} }_i = n \ 
\ket{ \downarrow_0 \cdots \downarrow_n \uparrow_{n+1} \downarrow_{n+2}
\cdots \downarrow_{N_P} }_i$. Notice that $[\bar{b}^{\dagger}_i,
\sum_{n=0}^{N_P} \sigma_z^{n,i}]=0$, which means that these operators
conserve the total $z$-component of the spin and, thus, always keep
states within the same subspace.

We can now write down the Hamiltonian in Eq. \ref{bosonhamilt}
in the spin-1/2 algebra as
\begin{equation}
\label{bosonhamilt2}
H=\sum\limits_{i,j=1}^N \alpha_{ij} \ \bar{b}^{\dagger}_i
\bar{b}^{\;}_j + \beta_{ij} \  \bar{n}_i \bar{n}_j  , 
\end{equation}
where the operators $\bar{b}^{\dagger}_i$ ($\bar{b}^{\;}_i$) are given
by Eq. \ref{bosonmap2} and $\bar{n}_i$ bt Eq. \ref{bosonmap2d}. In this
way, we are able to obtain physical properties of the bosonic system
(such as the  mean value of an observable, the mean value of the
evolution operator, etc.) in a QC made of qubits. It is important to
note that the  type of Hamiltonian given by Eq. \ref{bosonhamilt} is
not the only one that can be simulatable using the described method.
The only constraint is a fixed maximum number of bosons per site (or
mode).
\begin{figure}[hbt]
\begin{center}
\includegraphics[height=10.0cm]{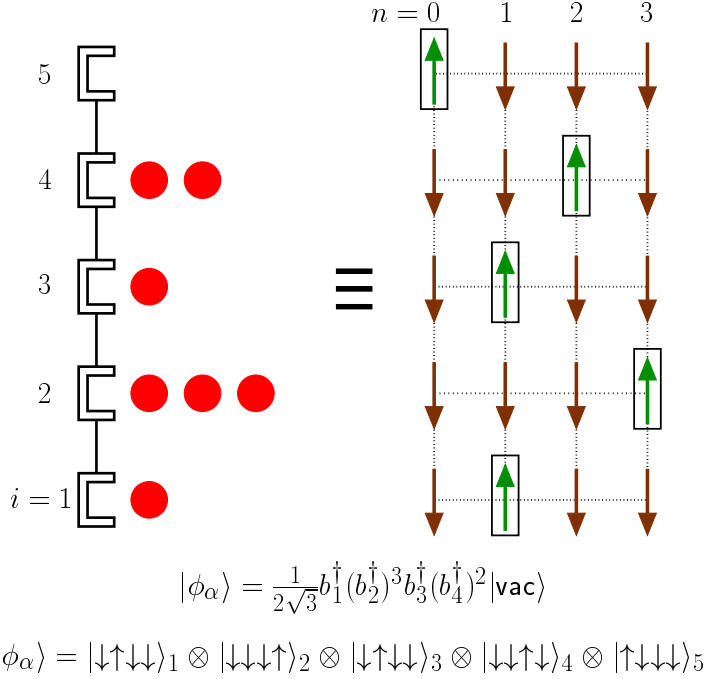}
\end{center}
\caption{Mapping of the bosonic state $\ket{\phi_\alpha}$, of a chain
of 5 sites and 7 bosons, into a spin-1/2 state (Eq. \ref{bosonmap}).}
\label{fig:2}
\end{figure}

\subsection{Preparation of the initial state}
\label{section4.1}
As in the fermionic case, the most general bosonic state of an $N$
sites quantum system with $N_P$ bosons can be written as a  linear
combination of product states like

\begin{equation}
\label{prodstate}
\ket{\phi_\alpha}= {\sf K} (b^\dagger_1)^{n_1}(b^\dagger_2)^{n_2} \cdots
(b^\dagger_N)^{n_N} \ \ket{\sf vac} ,
\end{equation}
where ${\sf K}$ is a normalization factor, $n_i$ is the number of 
bosons at site $i$ ($\sum\limits_{i=1}^N n_i = N_P$), and $\ket{\sf
vac}$ is the boson vacuum state (no particle state). Using the mapping
described in Eq. \ref{bosonmap}, we can write the vacuum state in the
spin language as  $\ket{\sf vac}=\ket{ \uparrow_0 \downarrow_1 \cdots
\downarrow_{N_P}}_1 \otimes \cdots \otimes \ket{ \uparrow_0
\downarrow_1 \cdots \downarrow_{N_P}}_N$  and  $\ket{\phi_\alpha} =
\ket{ \downarrow_0 \cdots \uparrow_{n_1} \cdots \downarrow_{N_P}}_1
\otimes \cdots \otimes \ket{ \downarrow_0 \cdots \uparrow_{n_N} \cdots
\downarrow_{N_P}}_N$ (see Fig. 2 for an example). Therefore, the
preparation of $\ket{\phi_\alpha}$ in a QC made of qubits is an easy
process: only $N$ spins are flipped from the fully polarized state,
where all spins are pointing down.

The preparation of a bosonic initial state of the form $\ket{\psi}=
\sum\limits_{\alpha=1}^L g_{\alpha} \ \ket{\phi_\alpha}$ is realized
as in the fermionic case. Again, we need to add $L$ ancillas (extra
qubits), perform controlled evolutions on their states, and finally
perform a measurement of an spin component \cite{ortiz2001}.

\subsection{Evolution of the initial state}
\label{section4.2}

The basic idea is to use the first order Trotter approximation  (see
the fermionic case) to  separate those terms of the Hamiltonian that
belong to the kinetic energy $K$, from the ones that belong to the
potential energy $V$ ($H=K+V$, $[K,V]\neq0$), i.e., 
\begin{equation}
\label{trotter2}
e^{i H \Delta t} \sim e^{i K \Delta t} e^{i V \Delta t}.
\end{equation}
\begin{figure}[hbt]
\begin{center}
\includegraphics[width=9.5cm]{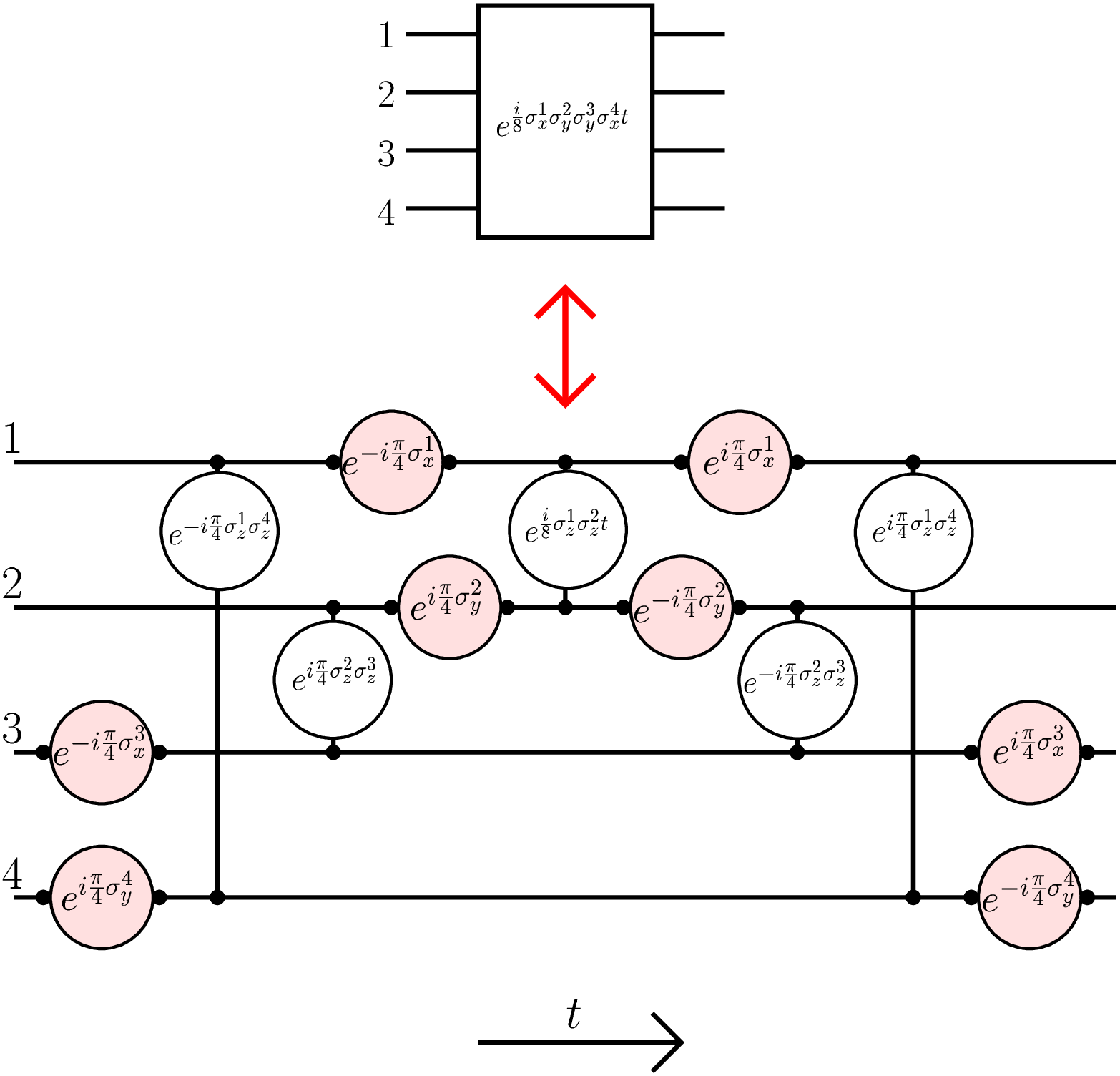}
\end{center}
\caption{Decomposition of the unitary operator $U(t)=e^{\frac{i}{8}
\sigma_x^1 \sigma_y^2 \sigma_y^3 \sigma_x^4 t}$ into single qubit
rotations and two qubits interactions. Time $t$ increases from left to
right.}
\label{fig:3}
\end{figure}

In general, $K$ is a sum of non commuting terms of the form
$b^{\dagger}_k b^{\;}_l + b^{\dagger}_l b^{\;}_k$, and we need to
perform another first order Trotter  approximation to decompose it into
elementary gates (in the spin language).  Then, a typical term $e^{i
(b^{\dagger}_i b^{\;}_j + b^{\dagger}_j b^{\;}_i)t}$ when mapped onto
the spin language (Eq. \ref{bosonmap2}) gives
\begin{eqnarray}
\label{decomp3}
\exp [ \frac{i t}{8} \sum\limits_{n,n'=0}^{N_P-1} 
\sqrt{(n+1)(n'+1)} \ [(\sigma_x^{n,i}\sigma_x^{n+1,i}
+\sigma_y^{n,i}\sigma_y^{n+1,i})(\sigma_x^{n',j}\sigma_x^{n'+1,j}
+\sigma_y^{n',j}\sigma_y^{n'+1,j}) \nonumber \\ 
+ (\sigma_x^{n,i}\sigma_y^{n+1,i}
-\sigma_y^{n,i}\sigma_x^{n+1,i})(\sigma_x^{n',j}\sigma_y^{n'+1,j}
-\sigma_y^{n',j}\sigma_x^{n'+1,j})]  ] ,
\end{eqnarray}
where $N_P$ is the number of bosons. The terms in the  exponent of Eq.
\ref{decomp3} commute with each other, so the decomposition into
elementary gates becomes straightforward. As an example (see Fig.3), 
we consider a
system of two sites with one boson. We need then $2(1+1)=4$ qubits for
the simulation, and  Eq. \ref{bosonmap2} implies that
$\bar{b}^\dagger_1 = \sigma_-^{0,1}\sigma_+^{1,1}$ and
$\bar{b}^\dagger_2 = \sigma_-^{0,2}\sigma_+^{1,2}$. Then, $e^{i
(b^\dagger_i b^{\;}_j  + b^\dagger_j b^{\;}_i)t}$ becomes
\begin{eqnarray}
\label{bosexamp}
\exp (\frac{it}{8} \sigma_x^{0,1}\sigma_x^{1,1}\sigma_x^{0,2}\sigma_x^{1,2})
\times
\exp (\frac{it}{8} \sigma_x^{0,1}\sigma_x^{1,1}\sigma_y^{0,2}\sigma_y^{1,2})
\times
\exp (\frac{it}{8} \sigma_y^{0,1}\sigma_y^{1,1}\sigma_x^{0,2}\sigma_x^{1,2})
\times
\exp (\frac{it}{8} \sigma_y^{0,1}\sigma_y^{1,1}\sigma_y^{0,2}\sigma_y^{1,2})
\\
\nonumber
\times
\exp (\frac{it}{8} \sigma_y^{0,1}\sigma_x^{1,1}\sigma_y^{0,2}\sigma_x^{1,2})
\times
\exp (-\frac{it}{8} \sigma_y^{0,1}\sigma_x^{1,1}\sigma_x^{0,2}\sigma_y^{1,2})
\times
\exp (-\frac{it}{8} \sigma_x^{0,1}\sigma_y^{1,1}\sigma_y^{0,2}\sigma_x^{1,2})
\times
\exp (\frac{it}{8} \sigma_x^{0,1}\sigma_y^{1,1}\sigma_x^{0,2}\sigma_y^{1,2}),
\end{eqnarray}
where the decomposition of each of the terms in Eq. \ref{bosexamp} in 
elementary gates can be done using the methods described in previous
works \cite{ortiz2001,somma2002}. In particular, in Fig. 3 we show
the decomposition of the term $\exp \left (\frac{i}{8}
\sigma_x^{0,1}\sigma_y^{1,1}\sigma_y^{0,2}\sigma_x^{1,2} t\right )$,
where the qubits were relabeled as $(n,i) \equiv n+2i -1$ (e.g.,
$(0,1)\rightarrow 1$).

On the other hand, it is important to mention that the number of
operations involved in the decomposition is not related to the distance
between the sites $i$ and $j$, as in the fermionic case.

\section{Measurement: Correlation functions and Energy spectra}
\label{section5}

In previous work \cite{ortiz2001,somma2002} we introduced an efficient
algorithm for the measurement of correlation functions in quantum
systems. The idea is to make an indirect measurement, that is, we
prepare an ancilla qubit (extra qubit) in a given initial state, then 
interact with the system whose properties one wants to measure, and
finally we measure some observable of the ancilla to obtain 
information about the system. Particularly, we could be interested in
the measurement of dynamical correlation functions of the form
\begin{equation}
\label{greenfunction}
G(t)= \langle \psi | T^{\dagger} A^\dagger_i T B_j \psi \rangle
\end{equation}
where $A_i$ and $B_j$ are unitary operators (any operator can be
decomposed in a unitary operator basis as $A=\sum\limits_i \alpha_i
A_i$, $B=\sum\limits_j \beta_j B_j$), $T=e^{-iHt}$ is the time
evolution operator of a time-independent Hamiltonian $H$, and
$\ket{\psi}$ is the state of the system whose correlations one wants to
determine. If we were interested in the evaluation of spatial
correlation functions, we would replace the evolution operator $T$ by
the space translation operator. In Fig. 4 we show the quantum algorithm
(quantum network) for the evaluation of $G(t)$. As explained before
\cite{ortiz2001,somma2002}, the initial state (ancilla plus system) has
to be prepared in the quantum state $\ket{+}_{\sf a} \otimes
\ket{\psi}$  (where ${\sf a}$ denotes the ancilla qubit and
$\ket{+}=\frac{\ket{0}+\ket{1}}{\sqrt{2}}$). Additionally, we have to
perform an evolution (unitary operation) in the following three steps:
i) a controlled evolution in the state $\ket{1}$ of the ancilla
$\mbox{C-B}= \ket{0} \langle 0 | \otimes I + \ket{1} \langle 1 |
\otimes B_j$, ii) a time evolution $T$, and iii) a controlled evolution
in the state $\ket{0}$ of the ancilla $\mbox{C-A}= \ket{0} \langle 0 |
\otimes A_i + \ket{1} \langle 1 | \otimes I$. Finally we measure the
observable $\langle 2\sigma_+^{\sf a} \rangle=  \langle \sigma_x^{\sf
a} +i \sigma_y^{\sf a}\rangle =G(t)$. 
\vspace*{-0.8cm}
\begin{figure}[hbt]
\begin{center}
\includegraphics[width=7.0cm]{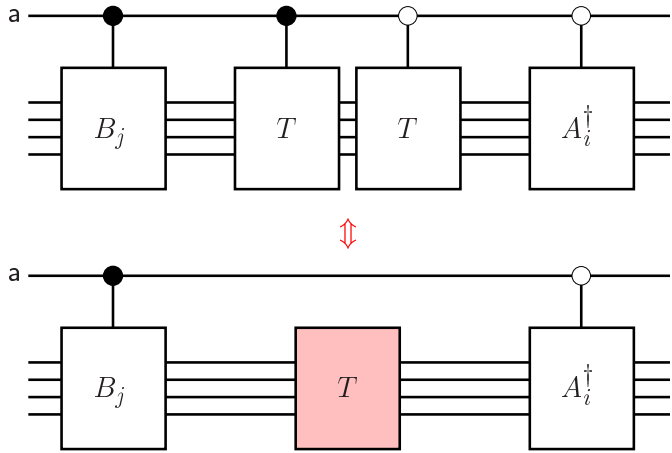}
\end{center}
\caption{Quantum network for the evaluation of $G(t)=
\langle \psi | T^{\dagger} A^\dagger_i T B_j \psi \rangle$.}
\label{fig:4}
\end{figure}

On the other hand, sometimes we are interested in obtaining the
spectrum (eigenvalues) of a given observable $\hat{Q}$ (i.e., an
hermitian operator). A quantum algorithm (network) for this purpose was
also given in previous work \cite{somma2002}. Again, the basic idea is
to perform an indirect measurement using an extra qubit (see Fig. 5).
Basically, we prepare the initial state (ancilla plus  system)
$\ket{+}_{\sf a} \otimes \ket{\phi}$, then apply the evolution  $e^{i
\hat{Q} \sigma_z^{\sf a} \frac{t}{2}}$, and finally measure  the
observable $\langle 2 \sigma_+^{\sf a}(t) \rangle = \langle \phi |
e^{-i \hat{Q} t} \phi \rangle$. Since the initial state of the system
can be written as a linear combination of eigenstates of $\hat{Q}$,
$\ket{\phi}=\sum \limits_{n=0}^L \gamma_n \  \ket{\psi_n}$, where
$\gamma_n$ are complex coefficients and $\ket{\psi_n}$ are eigenstates
of $\hat{Q}$ with eigenvalue $\lambda_n$, the classical Fourier
transform applied to the function of time $\langle 2\sigma_+^{\sf a}(t)
\rangle$ gives us $\lambda_n$
\begin{equation}
\label{spectrum}
\hat{F}(\lambda) = \sum\limits_{n=0}^L 2 \pi |\gamma_n|^2 \delta(\lambda - 
\lambda_n) .
\end{equation}
Without loss of generality, we can choose $\hat{Q}=H$, with $H$ some
particular Hamiltonian.
 
It is important to note that in order to obtain the different
eigenvalues of $\hat{Q}$, the overlap between the initial state and the
eigenstates of $\hat{Q}$ must be different from zero. One can use
different mean-field solutions of $\hat{Q}$ as initial states
$\ket{\phi}$ depending on the part of the spectrum one wants to
determine with higher accuracy.
\vspace*{-1.3cm}
\begin{figure}[hbt]
\begin{center}
\includegraphics[width=5.5cm]{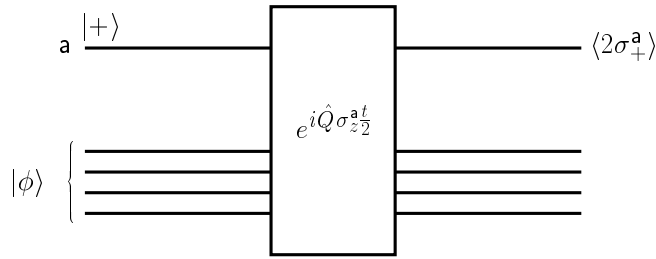}
\end{center}
\caption{Quantum network for the evaluation of the spectrum of an
observable $\hat{Q}$.}
\label{fig:5}
\end{figure}

\section{Algorithm efficiency and errors}
\label{section6}

An algorithm is considered efficient if the number of operations
involved scales polynomially with the system size, and if   the effort
required to make the error $\epsilon$ in the  measurement of a relevant
property smaller, scales polynomially with $1/\epsilon$.

While the evolution step involves a number of unitary operations that
scales polynomially with the system size (such is the case for the
Trotter  approximation) whenever the Hamiltonian $H$ is physical (e.g.,
is a sum of a number of terms that scales polynomially with the system
size),  the preparation of the initial state could be inefficient. Such
inefficiency would arise, for example, if  the state $\ket{\psi}$
defined in Eq. \ref{slater1} or Eq. \ref{prodstate} is a linear
combination of an  exponential number of states ($L\sim x^N$, with $N$
the number of sites in the system and $x$ a positive number). However,
if we assume that $\ket{\psi}$ is a finite combination of states ($L$
scales  polynomially with $N$), its preparation can be done
efficiently. (Any (Perelomov-Gilmore) generalized coherent state can be
prepared in a number of steps that scales polynomially with the 
number of generators of the respective algebra.) 
On the other hand, the measurement
process described in section \ref{section5} is always an efficient
step, since it only involves the  measurement of the spin of one qubit,
despite the number of qubits or sites $N$ of the quantum system.

Errors $\epsilon$ come from gate imperfections, the use of the Trotter
approximation in the evolution operator, and the statistics in
measuring the spin of the ancilla qubit (sections \ref{section3.2},
\ref{section4.2}, and \ref{section5}). A precise description and study
of the error sources can be found in previous work \cite{ortiz2001}.
The result is that the algorithms described here, for the simulation of
physical systems and processes, are efficient if the preparation of the
initial state is efficient, too.

\section{conclusions}
\label{section7}
We studied the implementation of quantum algorithms for the simulation
of an arbitrary quantum physical system on a QC made of qubits, making 
a distinction between systems that are governed by Pauli's exclusion
principle (fermions, hard-core bosons, anyons, spins, etc.), and
systems that are not (e.g, canonical bosons). For the first class of
quantum systems, we showed that a mapping between the corresponding
algebra of operators and the spin-1/2 algebra exists, since both have a
finite-dimensional representation. On the other hand, the operator
representation of quantum systems that are not governed by an
exclusion principle is infinite-dimensional, and an isomorphic mapping
to the spin-1/2 algebra is not possible. However, one can work with a
finite set of quantum states, setting a constraint, such as fixing the
number of bosons in the system. Then, the  representation of bosonic
operators becomes finite-dimensional, and we showed that we can write
down bosonic operators in the spin-1/2 language (Eq. \ref{bosonmap2}),
mapping bosonic states to spin-1/2 states (Eq. \ref{bosonmap}) .

We also showed how to perform quantum simulations in a QC made of
qubits (quantum networks), giving algorithms for the preparation of
the initial state, the evolution, and the measurement of a relevant
physical property, where in the most general case the unitary operations
have to be approximated (sections \ref{section3.2},\ref{section4.2}).

The mappings explained are efficient in the sense that we can perform
them in a number of operations that scales polynomially with the system
size. This implies that the evaluation of some correlation functions in
quantums states that can be prepared efficiently is also efficient,
showing an exponential speed-up of these algorithms with respect to
their classical simulation. However, these mappings are insufficient to
establish that quantum networks can simulate any physical problem
efficiently. As we mentioned in the introduction, 
this is the case for the determination of the spectrum of
the Hamiltonian in the two-dimensional Hubbard model \cite{somma2002},
where the signal-to-noise ratio decays exponentially with the system
size. 

Finally, in Fig. 6 a table displays the advantages of simulating some
known algorithms with a QC than with a CC, concluding that QCs behave
as efficient devices for some quantum simulations.

\begin{figure}[hbt]
\begin{center}
\includegraphics[width=3.5cm]{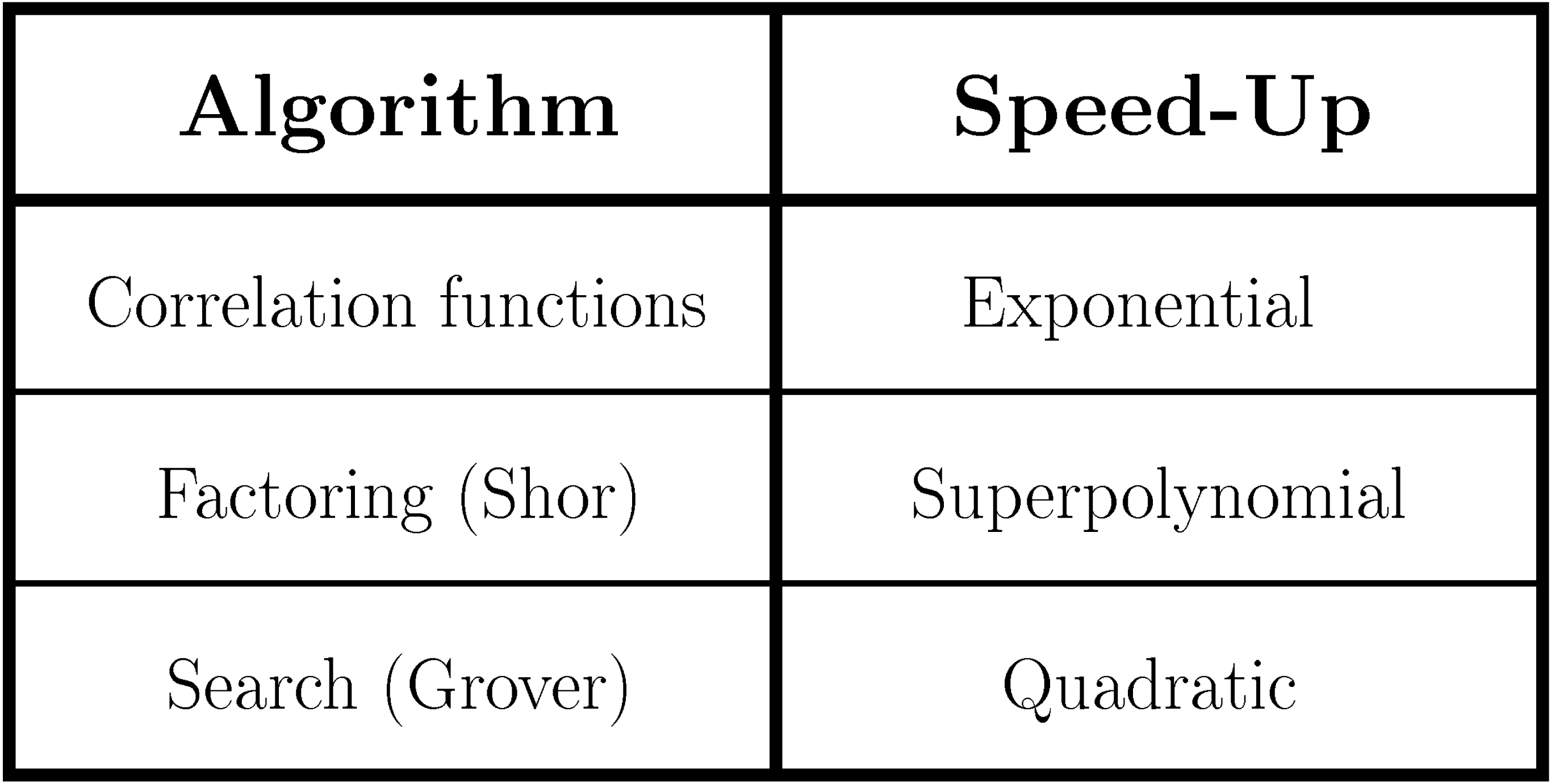}
\end{center}
\caption{Quantum vs. classical simulations. Speed-up refers to the gain
in speed of the quantum algorithms compared to the known classical
ones.}
\label{fig:6}
\end{figure}



\end{document}